\documentclass[aps,preprint,nofootinbib]{revtex4}
\usepackage{sublabel}
\usepackage{graphicx}
\usepackage{epsf}
\usepackage{wrapfig}
\usepackage{epsfig}
\def\Vec#1{\mbox{\boldmath $#1$}}
\def\bra#1{\langle#1\,|}
\def\ket#1{|\,#1\rangle}

\def\beq{\begin{equation}}
\def\eeq{\end{equation}}

\def\beqy{\begin{eqnarray}}
\def\eeqy{\end{eqnarray}}
\def\ie{i.e.}

\def\ma{\hat{A}}
\def\mh{\hat{\mathcal{O}}}

\begin{document}
\vskip 2mm \date{\today}\vskip 2mm
\title{A Realistic Calculation of the Effects of Nucleon-Nucleon
  Correlations in High-Energy Scattering Processes Off Nuclei}
\author{M. Alvioli}
\author{C. Ciofi degli Atti}
\author{I. Marchino}
\address{Department of Physics, University of Perugia and
      Istituto Nazionale di Fisica Nucleare, Sezione di Perugia,
      Via A. Pascoli, I-06123, Perugia, Italy}
\author{H. Morita}
\address{Sapporo Gakuin University, Bunkyo-dai 11, Ebetsu 069-8555,
Hokkaido, Japan}
\vskip 1cm
\begin{abstract}
\noindent A new linked cluster expansion for the calculation of ground state observables 
of complex nuclei with realistic interactions has been developed \cite{alv01,alv02,alv03};
using the $V8^\prime$ \cite{pud01} potential  the ground state energy, density and momentum
distribution
of complex nuclei have been calculated and found to be in good agreement with the results of
\cite{fab01}, obtained within the Fermi Hyper Netted Chain, and Variational Monte Carlo
\cite{pie01} approaches. Using the same cluster expansion, with wave function and correlations
parameters fixed from the calculation of the ground-state observables, the semi-inclusive
reaction of type $A(e, e^\prime p)X$ has been calculated taking final state interaction  
effects into account within a Glauber type calculation as in Ref. \cite{cio01}; the 
comparison between the resulting distorted and
undistorted momentum distributions provides an estimate of the transparency of the nuclear
medium to the propagation of the hit proton. The effect of color transparency has also 
been considered within the approach of \cite{bra01,mor01}; it is shown that at high values of
$Q^2$ finite  formation time effects strongly reduce the final state interaction, consistently
with the idea  of a reduced interaction of the hadron produced inside the nucleus \cite{bro01}.
The total neutron-nucleus  cross section at high energies has also been calculated \cite{alv04}
by  considering the effects of nucleon-nucleon correlations, which are found to increase the
cross section by about 10\% in disagreement with the experimental data. The inclusion of
inelastic shadowing effects of Refs. \cite{gri01,jen01} decreases back the cross section,
leading  to a good agreement between experimental data and  theoretical calculations.
\end{abstract}
\maketitle
\section{Introduction}

The knowledge of the nuclear wave function, in particular its most interesting
and poorly known  part, {\it viz} the correlated one, 
is not only a prerequisite for understanding the details of bound hadronic systems,
but is becoming at present a necessary ingredient for a correct description of
medium and high energy scattering processes off nuclear targets; these in fact
represent a current way of investigating short range effects in nuclei as well
as those QCD effects (e.g. color transparency,  hadronization, dense
 hadronic matter, etc)
which manifest themselves in the nuclear medium.
The necessity to treat nuclear effects in medium and high energy scattering 
 within a realistic many body description, 
becomes therefore
clear. The problem is not trivial, for one has first to solve the many body problem
and then has to find a way to apply it to scattering processes. 
The  difficulty mainly arises because even if  a reliable and manageable 
many-body description of the ground state is developed, the
problem remains of the calculation of the final state.
In the case of  \textit{few-body systems}, a consistent treatment of Initial
State Correlations (ISC) and final state interaction (FSI) is nowadays possible
at low energies by solving the Schr\"odinger equation for the bound and continuum states
but, at high energies, the  Schr\"odinger approach  becomes impractical and other methods
have to be employed.
In the case of \textit{complex nuclei}, much remains to be done, also in view that
the results of very sophisticated calculations
($e.g.$\,\, the variational Monte Carlo ones \cite{pie01}), show that 
the wave function which minimizes the expectation value of the Hamiltonian
provides a very poor nuclear density; moreover, the structure of the best trial 
wave function is so complicated, that its use in the 
calculation of various processes at intermediate and high energies
appears to be not easy task. It is for this reason that the evaluation of  nuclear
effects in medium and high energy scattering processes 
is usually  carried out within simplified models of nuclear structure. 
As a matter of fact, initial state  correlations ($ISC$)  are often  introduced
by a procedure which has little to recommend itself, namely the expectation 
value of the transition operator is evaluated with shell model (SM) uncorrelated wave functions
and  the final state two-nucleon SM wave function is replaced
by a phenomenological correlated wave function; to date,  a consistent treatment of
both ISC and FSI in intermediate and high energy scattering off complex nuclei
is far from being completed, so that a quantitative and unambiguous evaluation
of  the role played ISC is still lacking.
For such a reason we have undertaken the calculation of the ground-state
properties (energies, densities and momentum distributions) of complex nuclei within
a framework which can be easily generalized to the treatment of various scattering
processes, keeping  the basic features of ISC as predicted by the structure of
realistic Nucleon-Nucleon (NN) interactions.
This paper is organized as follows; in Section \ref{sectionCLUSTER} the cluster
expansion method is introduced, and the relevant ground state properties of $^{16}O$
and $^{40}Ca$ calculated. In Section \ref{sectionEEP} the semi-inclusive $A(e,e^\prime p)X$
reaction off complex nuclei targets is considered, and the FSI is calculated within the
Glauber and Finite Formation Time pictures, taking advantage of the wave functions
obtained in Section \ref{sectionCLUSTER}. In Section \ref{sectionXSECT} the total
neutron-nucleus cross section is calculated with the same wave functions, and the
result of taking into account correlations and inelastic shadowing effects are
discussed.
\section{The cluster expansion}\label{sectionCLUSTER}

We write the nuclear Hamiltonian in the usual form, \ie:
\beq
\label{hamilt}
\hat{H}\,=\,\hat{T}\,+\,\hat{V}\,
        =\,-\frac{\hbar^2}{2\,M_N}\,\sum^A_{i=1}\,\nabla^2_i\,
                \,+\,\sum_{i<j}\,\hat{v}_2(\Vec{x}_i,\Vec{x}_j)\;,
\eeq
where the vector $\Vec{x}$ denotes the set of nucleonic degrees of freedom,
and the two-body potential $\sum_{i<j}\,\hat{v}_2(\Vec{x}_i,\Vec{x}_j)$
is of the form
\beq
\label{defpotere}
\hat{V}\,=\,\sum_{n=1}^{N}\,v^{(n)}(r_{ij})\,\mh^{(n)}_{ij}\,,
\eeq
where $r_{ij}=|\Vec{r}_i-\Vec{r}_j|$ is the relative distance of
nucleons $i$ and $j$,  and $n$, ranging up to $N=18$, labels
the state-dependent operator $\mh^{(n)}_{ij}$:
\beq
\label{newuno}
\mh^{(n)}_{ij}\,=\,\left[1\,,\,\Vec{\sigma}_i\cdot\Vec{\sigma}_j\,
        ,\,\hat{S}_{ij}\,,\,\left(\Vec{S}\cdot\Vec{L}\right)_{ij}\,,
        \,L^2,..
        \right]\,\otimes\,\left[1\,,\,\Vec{\tau}_i\cdot\Vec{\tau}_j\right]\,.
\eeq
The evaluation of the expectation value of the nuclear Hamiltonian (\ref{hamilt})
is object of intensive activity which, in the last few years \cite{transp}, has produced
considerable results; nevertheless, the level of complexity of the obtained wave functions
is such
that they cannot be used in scattering problems with reasonable ease.
Our goal is to present a more economical,   but effective
method for the calculation of the expectation value of any quantum
mechanical operator $\ma$
\beq
\label{omedio1}
\langle\,\ma\,\rangle\,=\,\frac{\bra{\psi_o}\,\ma\,
        \ket{\psi_o}}{\bra{\psi_o}\psi_o\rangle}\,;
\eeq
with $\psi_o$ having the following structure
\beq
\label{psi1}
\psi_o(\Vec{x}_1,...,\Vec{x}_A)\,=\,\hat{F}(\Vec{x}_1,...,\Vec{x}_A)\,
        \phi_o(\Vec{x}_1,...,\Vec{x}_A)\,,
\eeq
where $\phi_o$ is a Shell-Model (SM),  mean-field wave function, and $\hat{F}$ is 
a symmetrized \textit{correlation operator}, which generates correlations
into the mean field wave function. According to the two-body interaction of Eq.
(\ref{defpotere}), the correlation operator is cast in the following form:
\beq
\label{newdue}
\hat{F}(\Vec{x}_1,\Vec{x}_2\,...\,\Vec{x}_A)\,=\,\hat{S}\,\prod^A_{i<j}\,\hat{f}(r_{ij})
\eeq
with
\beq
\hat{f}(r_{ij})=\sum_{n=1}^{N}\,\hat{f}^{(n)}(r_{ij})\;\;\hspace{2cm}
        \hat{f}^{(n)}(r_{ij})=f^{(n)}(r_{ij})\,\hat{O}^{(n)}_{ij}\,.
\label{corrop1}
\eeq
In the present paper we are going to introduce a cluster expansion technique
in order to evaluate Eq. (\ref{omedio1}); the solution can be found by applying
the variational principle, with the variational parameters contained both in the
correlation functions and in the mean field single particle wave functions.
The expectation value $\langle \hat{A} \rangle$ defined in (\ref{omedio1}) can
be expanded in the framework of the cluster expansion developed in Refs.
\cite{alv01,alv02,alv03} and, at first order, it reads as follows
\beq
\label{seventeen}
\langle\hat A\rangle_1=\bra{\phi_o}\sum_{i<j}\left(\hat{f}_{ij}\hat A
        \hat{f}_{ij}-\hat{A}\right)\ket{\phi_o}
      -\langle \hat A\rangle_o\bra{\phi_o}\sum_{i<j}
      \left(\hat{f}_{ij}
        \hat{f}_{ij}-1\right)\ket{\phi_o}\,,
\eeq
where $\langle \hat A\rangle_o$ is given by $\bra{\phi_o} \hat{A}\ket{\phi_o}$
The $2nd$ order term can straightforwardly be obtained by the same technique
used to derive Eq. (\ref{seventeen})
Given the two-body interaction as in Eq. (\ref{defpotere}), the expectation
value  of the Hamiltonian can be written in the following way:
\beqy
\label{hmatrix}
E_o&=&-\,\frac{\hbar^2}{2\,M_N}\,\int d\Vec{r}_1\left[\nabla_1\,\cdot\,\nabla_{1^\prime}\,
        \,\rho^{(1)}(\Vec{r}_1,\Vec{r}_1^\prime)\right]_{\Vec{r}_1=\Vec{r}_1^\prime}
        \,+\nonumber\\&&\hspace{2cm}+\,\sum_n\;\int\;d\Vec{r}_1 d
        \Vec{r}_2\;v^{(n)}(r_{12})\rho^{(2)}_{(n)}(\Vec{r}_1,\Vec{r}_2)\,,
\eeqy
where $\rho^{(1)}(\Vec{r}_1,\Vec{r}_1^\prime)$ and 
$\rho^{(2)}_{(n)}(\Vec{r}_1,\Vec{r}_2)$ are the non-diagonal One Body Density Matrix
(OBDM) and the (spin and isospin dependent-; see Ref. \cite{alv01}) Two Body Density
(TBD) matrices, respectively These can be calculated from the ground state wave
function by inserting in Eq. (\ref{omedio1}) the corresponding operators.
The knowledge of the OBDM and TBD matrices allows one to calculate, besides the
ground-state energy, other relevant quantities like \textit{e.g.} the density distribution:
\beq\label{obdrr}
\rho(\Vec{r})\,=\,\rho^{(1)}(\Vec{r}_1=\Vec{r}_1^\prime\equiv\Vec{r})\,,
\eeq
and the nucleon momentum distribution, defined as:
\beq
\label{defmomdis}
n(\Vec{k})\,=\,\frac{1}{(2\pi)^3}\,\int d\Vec{r}_1 d\Vec{r}^\prime_1
        \,e^{i\,\Vec{k}\cdot(\Vec{r}_1-\Vec{r}^\prime_1)}\,\rho(\Vec{r}_1,
        \Vec{r}^\prime_1)\,.
\eeq
The results of calculation of the ground state energy, the charge density and
the two-body density and momentum distribution using the realistic $V8^\prime$
interaction \cite{pud01} is discussed in detail in Ref. \cite{alv01,alv02,alv03}.
\begin{figure}[!htp]
\centerline{\epsfig{file=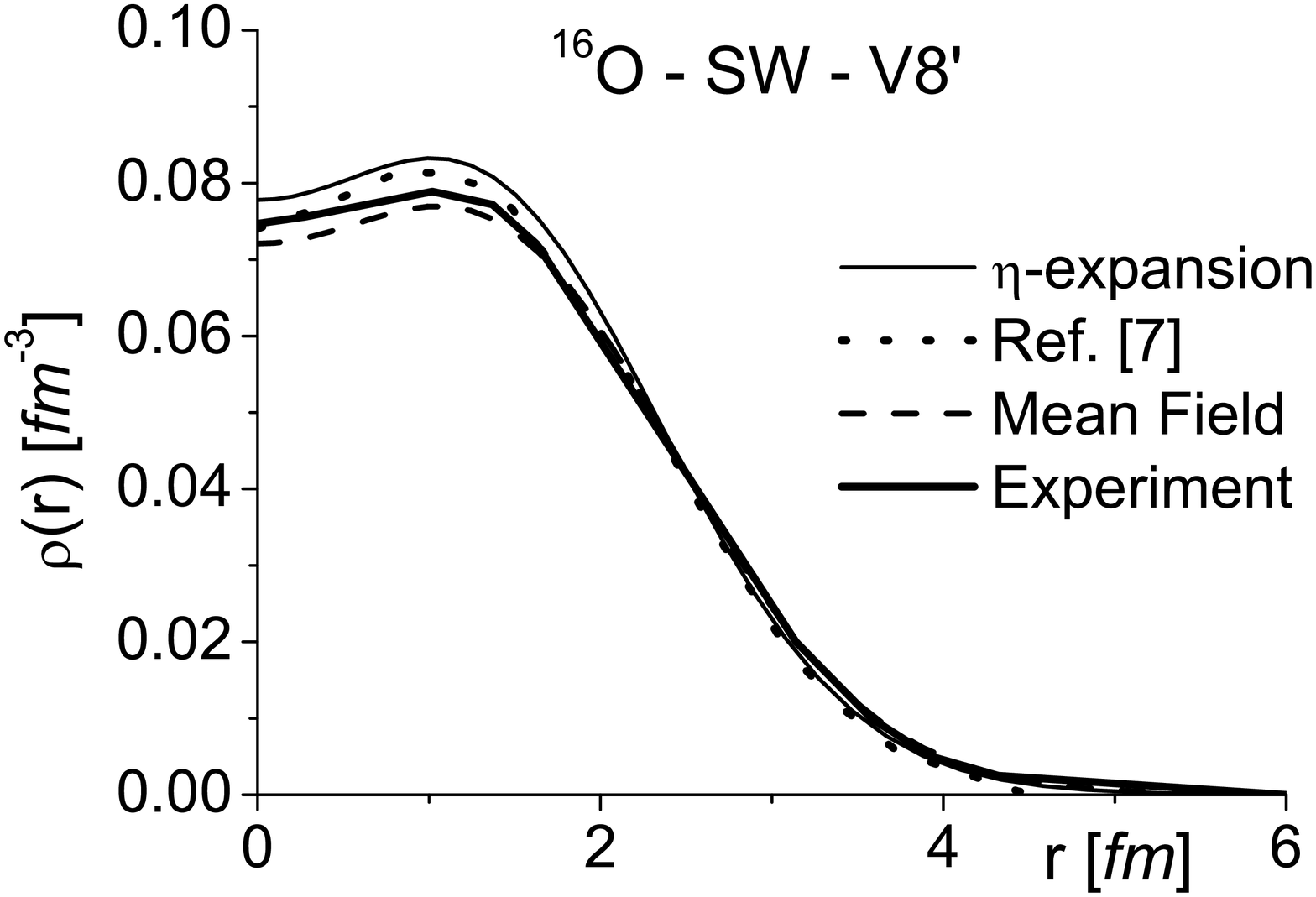,width=80mm}
   \epsfig{file=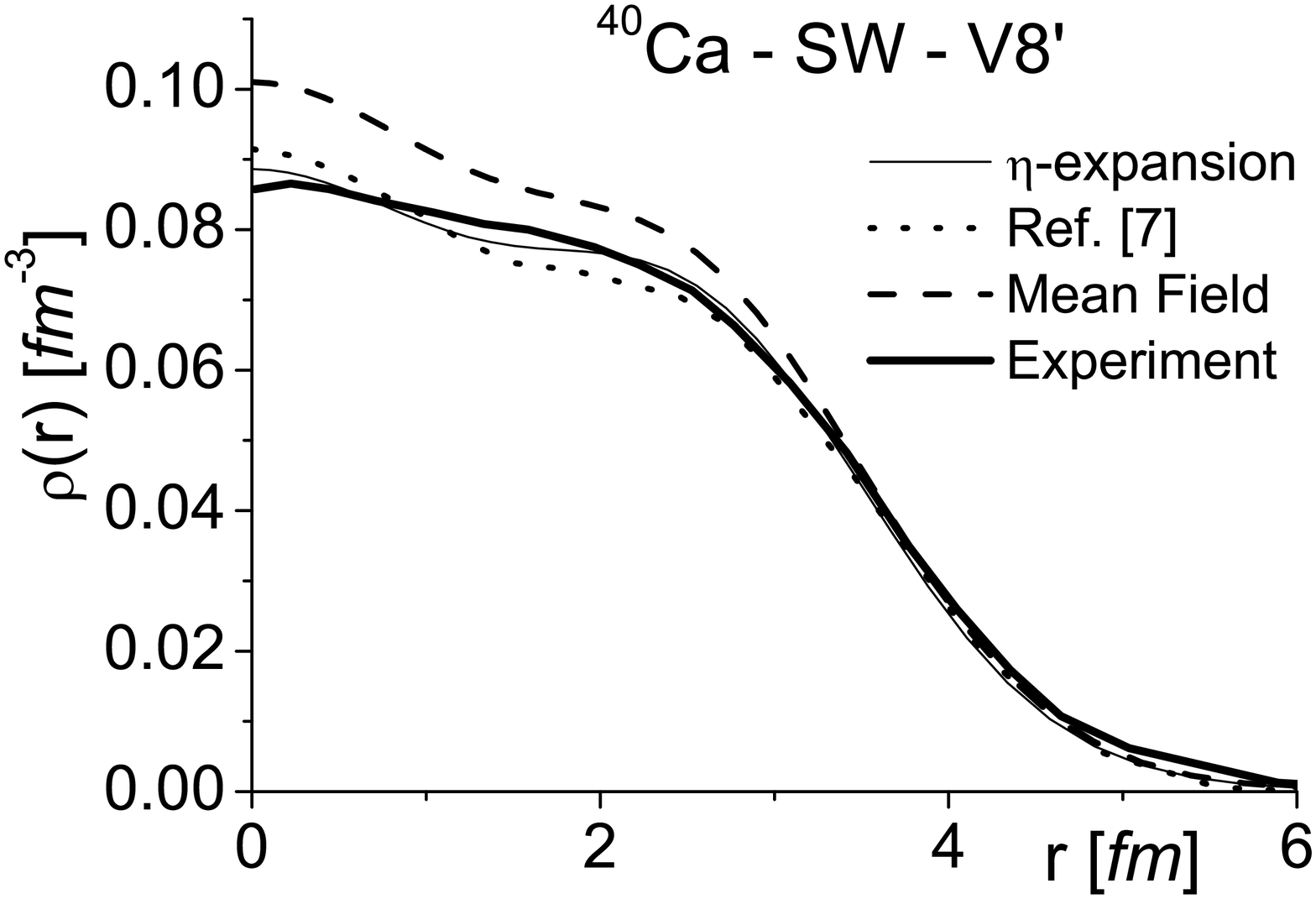,width=80mm}}
\caption{The charge densities of $^{16}O$ (\textit{left}) and $^{40}Ca$
(\textit{right}) calculated within the cluster expansion compared with the
experimental data. The mean field result and the results of the FHNC calculation
of Ref. \cite{fab01} are also shown.}\label{fig1}
\end{figure}
In Fig. \ref{fig3} we show the effect of correlations on the charge density and
the two-body density of $^{16}O$. In the case of charge density, we have shown the
three contributions which arises from the first-order cluster expansion, \ie, the
mean field, \textit{hole} and \textit{spectator} contributions, which are shown in
Fig. \ref{fig17} in a diagrammatic picture. In the case of two-body density, we
compare the mean field result with the ones obtained taking into account the only
central correlation ($f_1$ \textit{approximation}) and the first six correlations
($N = 6$ in Eq. (\ref{corrop1}); $f_6$ \textit{approximation}).
\begin{figure}[!hbp]
\centerline{\epsfig{file=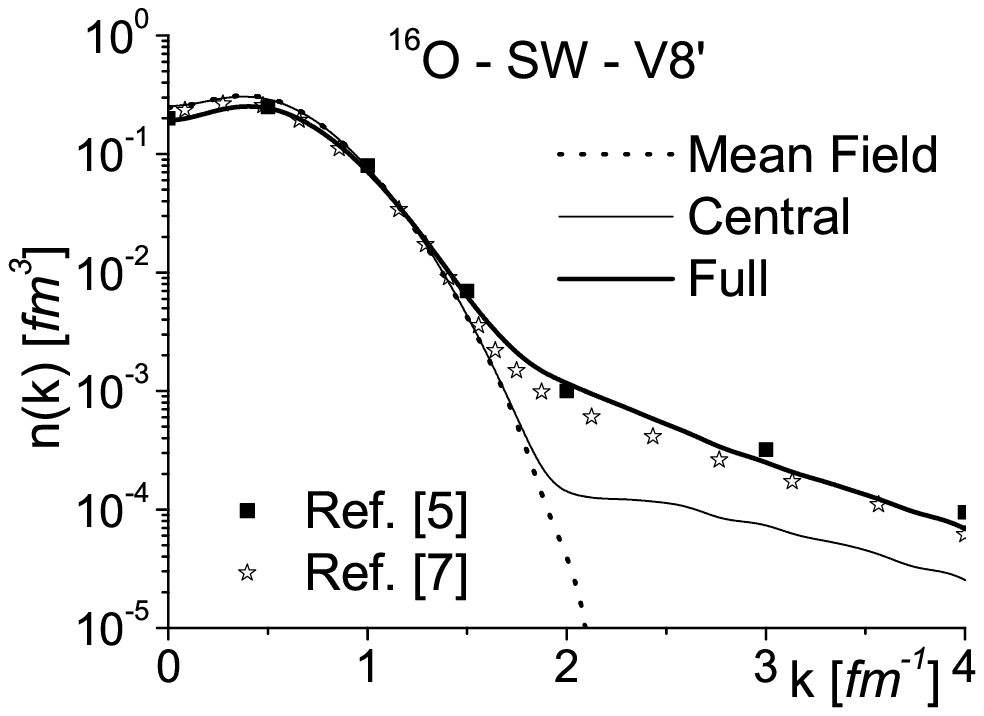,width=80mm}
   \epsfig{file=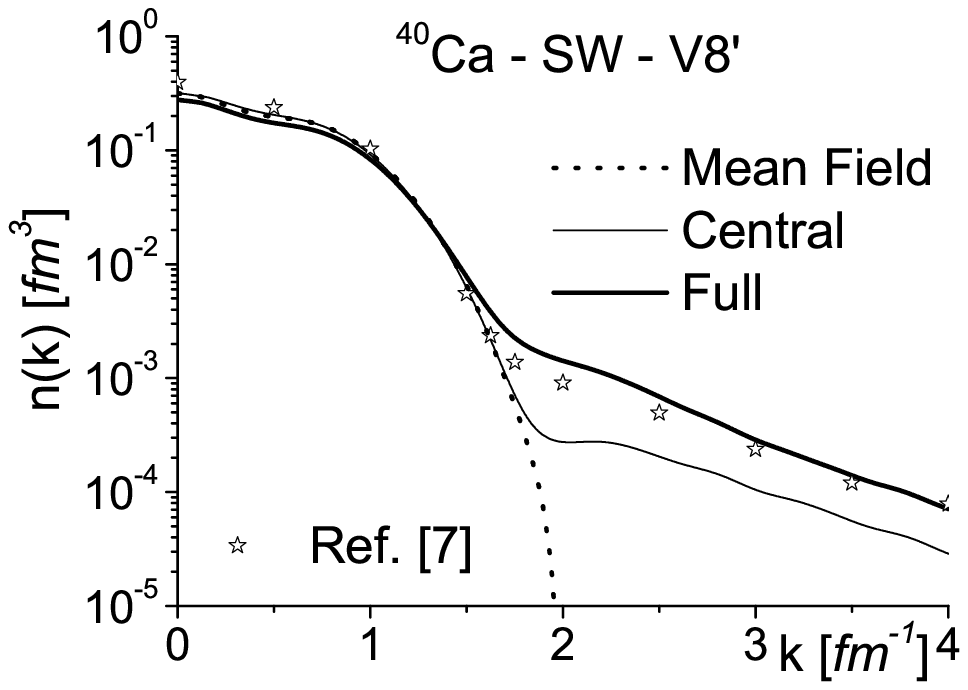,width=80mm}}
\caption{The momentum distributions of $^{16}O$ (\textit{left}) and $^{40}Ca$
(\textit{right}) calculated within the cluster expansion, compared with the
mean field result and with the FHNC approach of Ref. \cite{fab01} and the VMC
approach of Ref. \cite{pie01}.}\label{fig2}
\end{figure}
\begin{figure}[!htp]
\centerline{\epsfig{file=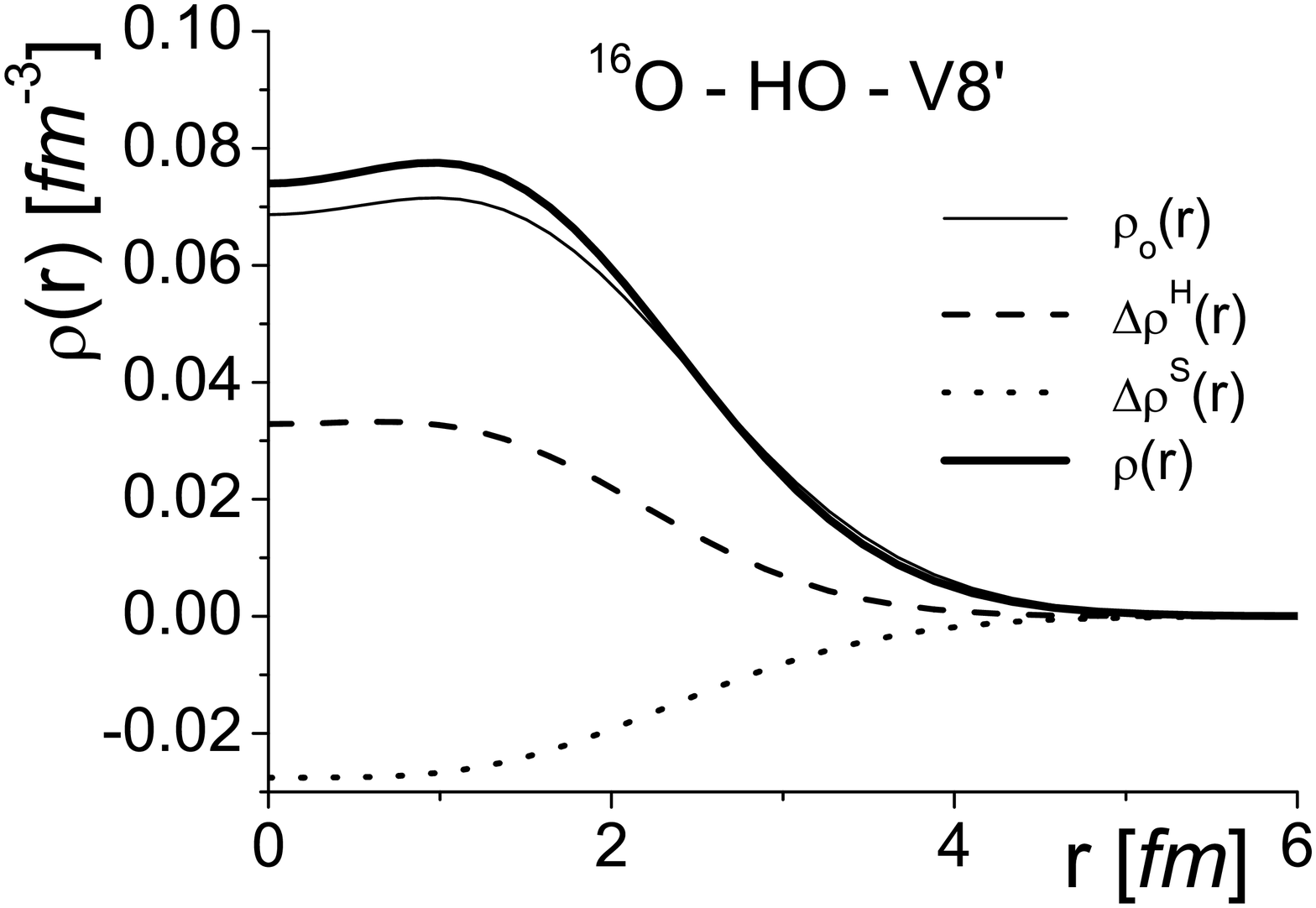,width=80mm}
   \epsfig{file=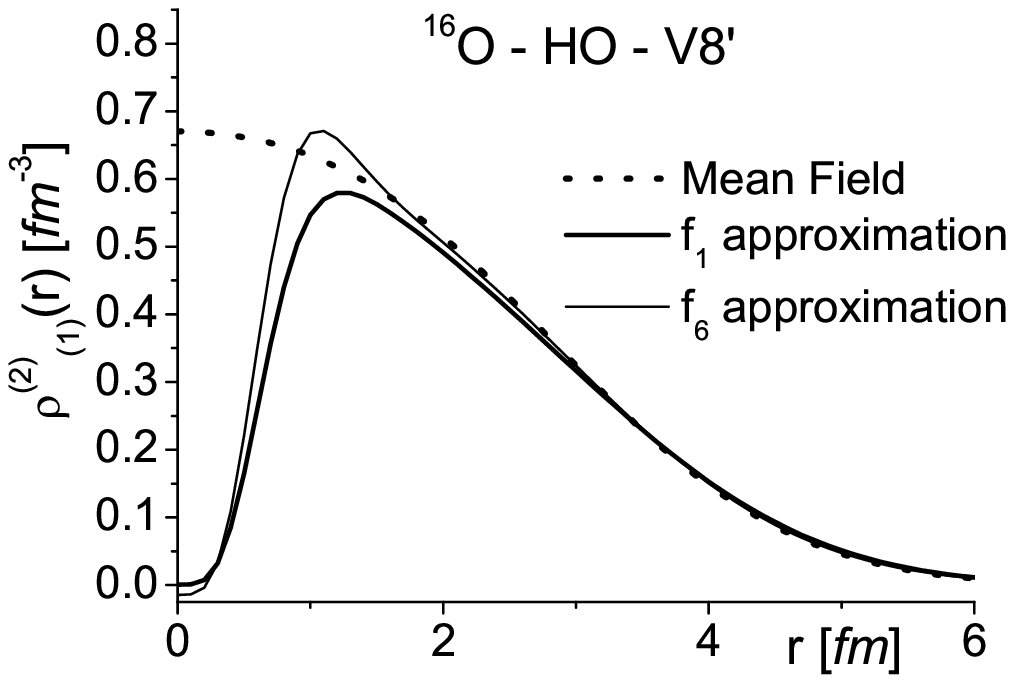,width=80mm}}
\caption{Effect of correlations versus the non-correlated result for 
the charge density (\textit{left}) and the two-body density (\textit{right}) 
of $^{16}O$ calculated within the cluster expansion.}\label{fig3}
\end{figure}
\begin{figure}[!hbp]
\centerline{\textit{a)}\hskip 1cm
      \epsfysize=0.9cm\epsfbox{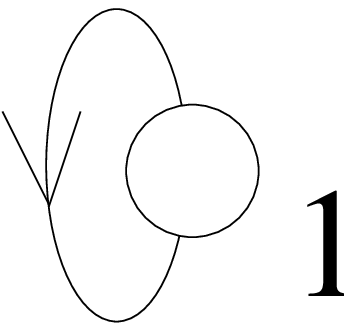}}
\vskip 0.8cm
\centerline{\textit{b)}\hskip 1cm
      \epsfysize=1.cm\epsfbox{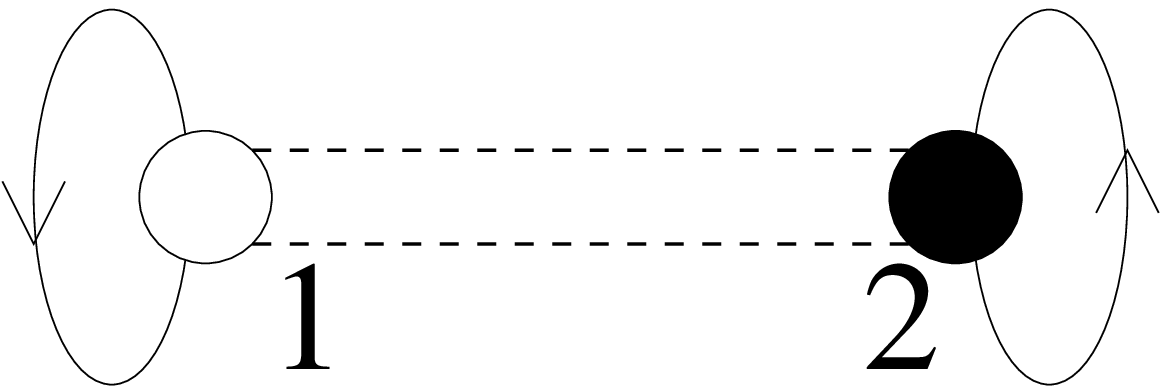}
      \hskip 2.5cm
      \epsfysize=1.1cm\epsfbox{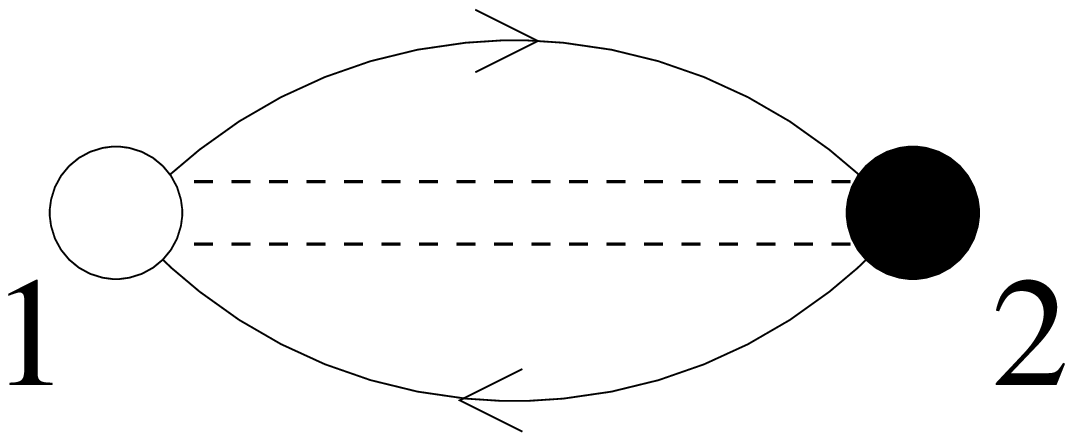}}
\vskip 1cm
\centerline{\textit{c)}\hskip 1cm
      \epsfysize=2.5cm\epsfbox{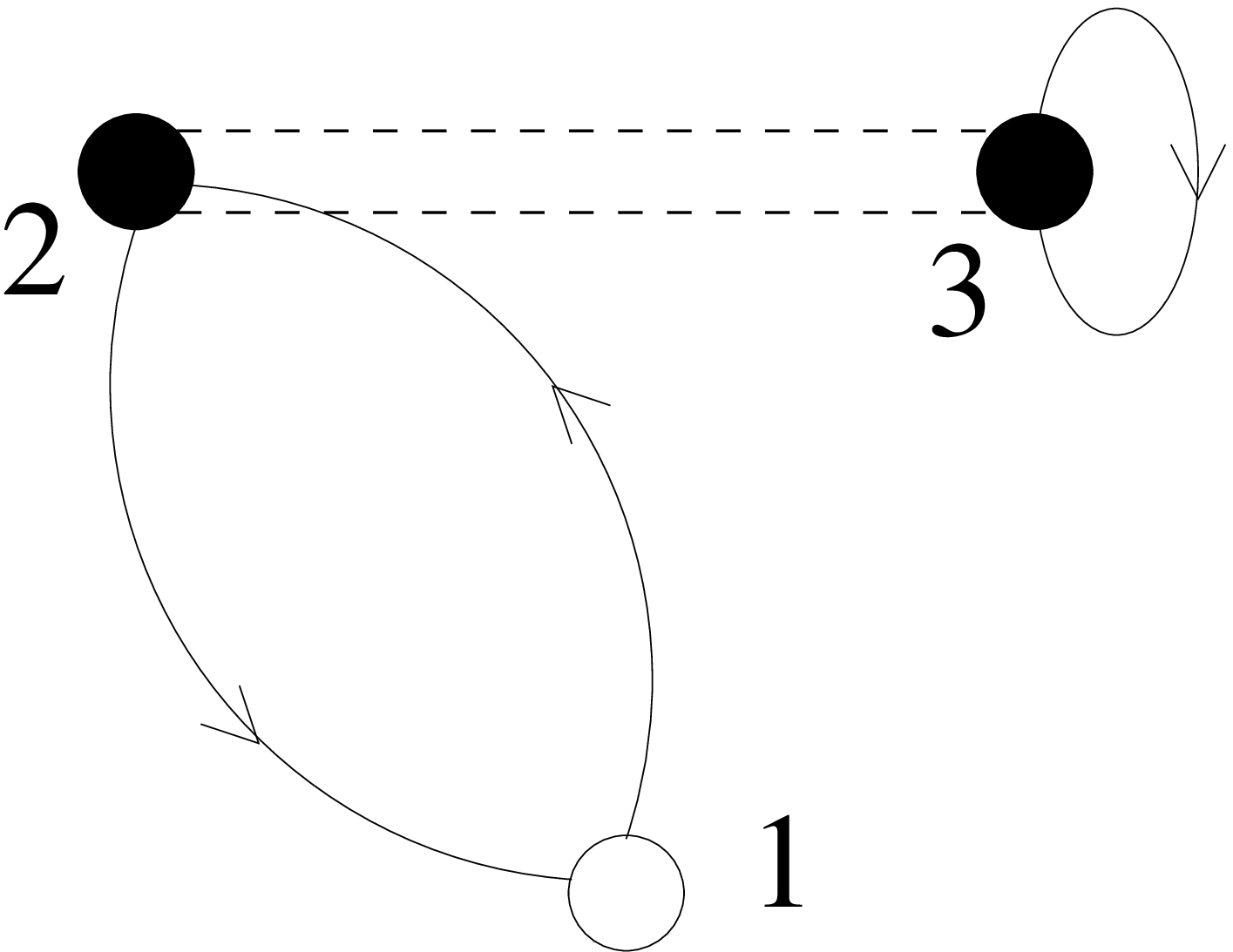}
      \hskip 3cm
      \epsfysize=2.5cm\epsfbox{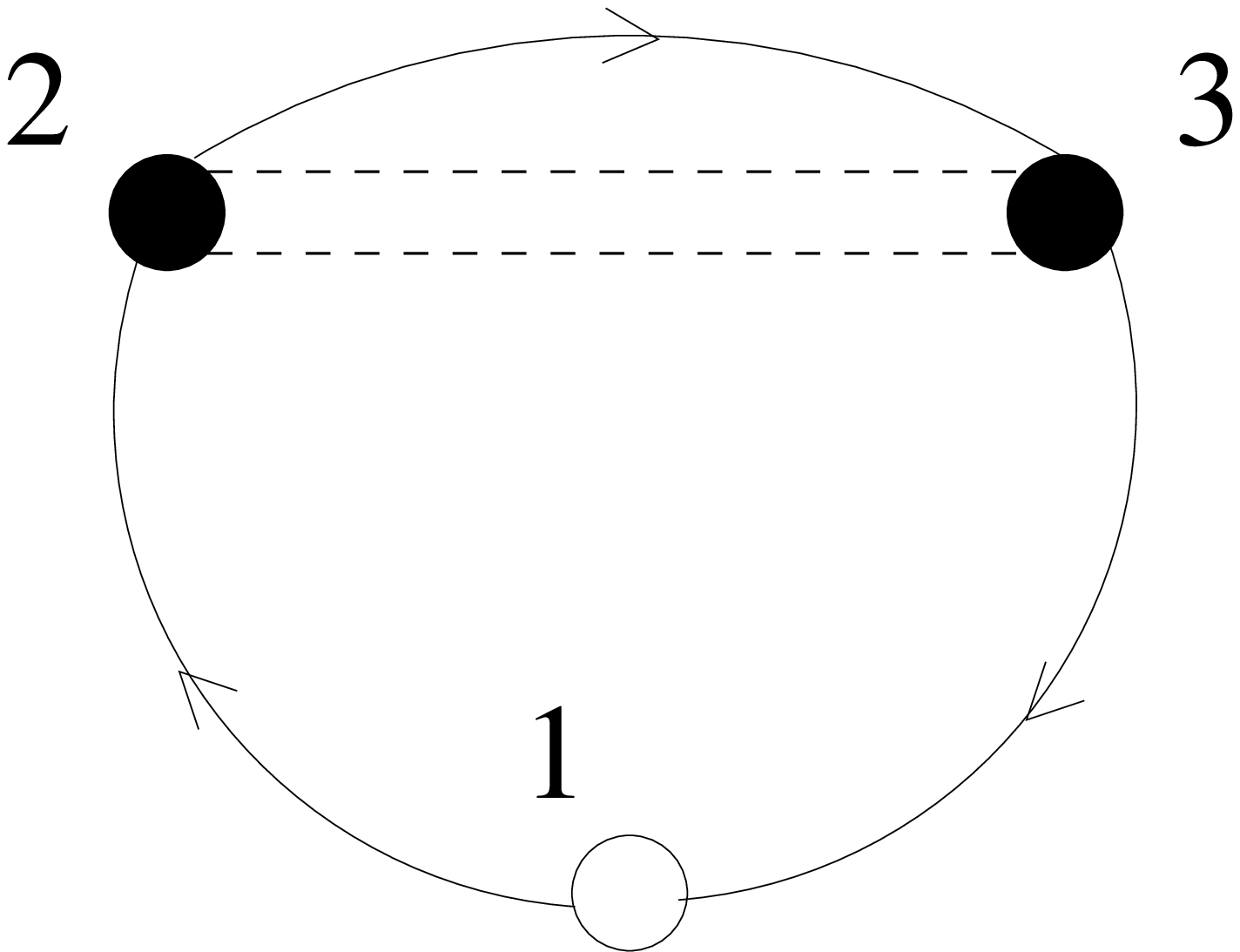}}
\caption{Diagrammatic representation of the one body mixed density matrix 
      $\rho(\Vec{r}_1,\Vec{r}^{\prime}_1)$ in the lowest order of the $\eta$-expansion
      (Eq. (\ref{obdrr})). \textit{a)} is the  shell model contribution and  $\textit{b)}$
      and $\textit{c)}$ the $hole$ and {\it spectator} contributions, respectively.
      The direct and exchange contributions are shown on the left and right sides
      of the Figure, respectively.}\label{fig17}
\vskip -0.1cm
\end{figure}

\section{Applications. I - the semi-inlcusive process $A(e,e^\prime p)X$}
\label{sectionEEP}

\begin{figure}[!htp]
\centerline{\epsfig{file=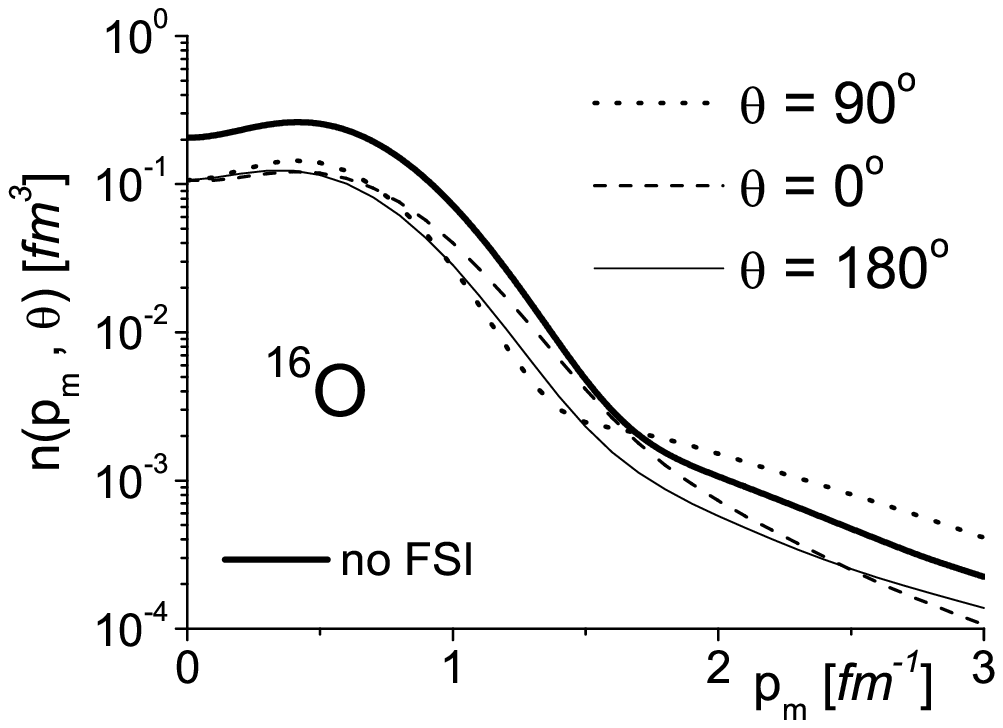,width=80mm}
   \epsfig{file=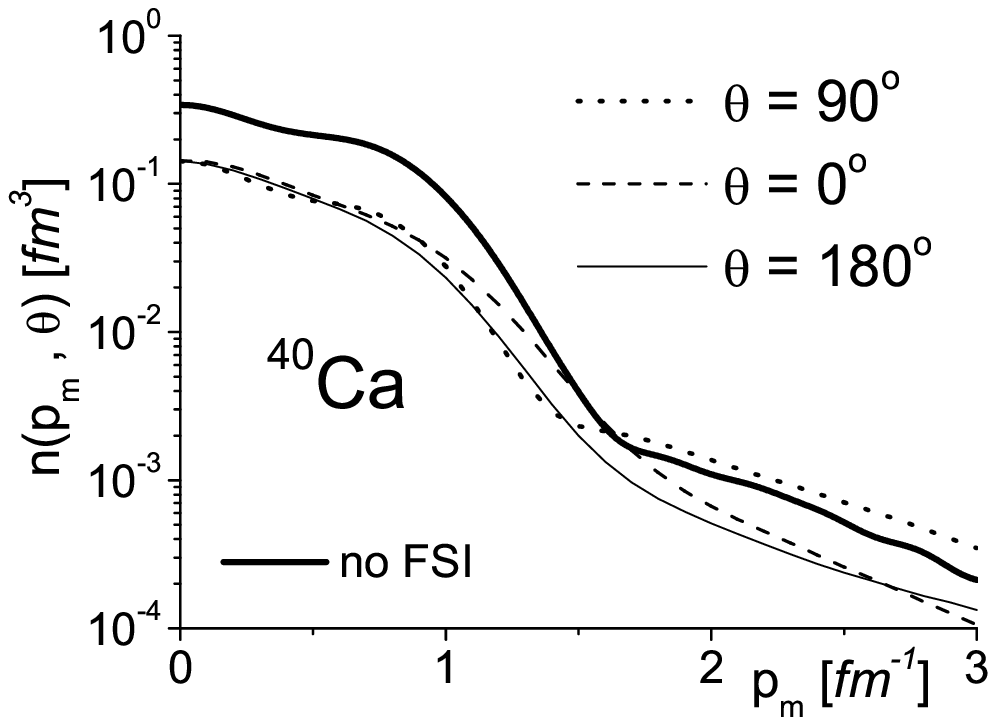,width=80mm}}
\caption{The distorted momentum distribution, $n_D({\bf p}_m)=n_D(p_m, \theta)$ 
      ($\theta=\widehat{{\bf q}{\bf p}}_m$),  of $^{16}O$  and $^{40}Ca$,
      obtained  by Eq. (\ref{nd}).}\label{fig4}
\end{figure}

Using the cluster expansion developed in in the Section \ref{sectionCLUSTER},
we have calculated
the semi-inclusive $A(e,e'p)X$ process in which an electron with 4-momentum
$k_1\equiv\{{\bf k}_1,i\epsilon_1\}$, is scattered off a nucleus with 4-momentum
$P_A\equiv\{{\bf 0},iM_A\}$ to a state $k_2\equiv\{{\bf k}_2,i\epsilon_2\}$ 
and is detected in coincidence with a proton $p$ with 4-momentum 
$p\equiv\{{\bf p},iE_p\}$; the final $(A-1)$ nuclear system with 
4-momentum $P_X\equiv\{{\bf P}_X,iE_X\}$ is undetected. The cross section 
for the exclusive process $A(e,e'p)B$  can be written as follows
\begin{equation}
\frac{d\sigma}{dQ^2 d\nu d{\bf p}}=K\sigma_{ep}P_D(E_m,{\bf p}_m)
\label{sezione}
\end{equation}
where $K$ is a kinematical factor, $\sigma_{ep}$ the off-shell electron-nucleon 
cross section,  and $Q^2=|{\bf q}|^2-\nu^2$ the four momentum transfer. The 
quantity $P_D(E_m,{\bf p}_m)$ is the distorted nucleon spectral function which 
depends upon the observable \textit{missing momentum} ${\bf p}_m={\bf q}-{\bf p}$
(${\bf p}_m=-{\bf k}$ when the FSI is absent)
and \textit{missing energy} $E_m=\nu-T_p -T_{A-1}$. 
In the semi-inclusive $A(e,e'p)X$ process,   the cross section (\ref{sezione}) is
integrated over the missing energy $E_m$, at fixed value of ${\bf p}_m$ and 
becomes directly proportional to the \textit{distorted} 
momentum distribution
\begin{equation}
n_D({\bf p}_m)={(2 \pi)^{-3}} \int e^{i {\bf p}_m\,\cdot\,({\bf r}_1 -{\bf r}_1')}
      \rho_D ({\bf r}_1,{\bf r}_1')\, d{\bf r}_1 d{\bf r}_1'
\label{nd}
\end{equation}
where 
\begin{eqnarray}
\rho_D ({\bf r}_1,{\bf r}_1')= \frac {\langle\psi_o\,S^{\dagger}
\,\hat{\rho}({\bf r}_1,{\bf r}_1')\,S'\,{\psi_o}^\prime\rangle}{\langle\psi_o
      \,\psi_o\rangle}
\label{rodi}
\end{eqnarray}
is the  distorted one-body mixed density matrix, $S$ is the S-matrix describing FSI
and\beq
\label{opuno}
\hat{\rho}_1(\Vec{\tilde{r}}_1,\Vec{\tilde{r}}_1^\prime)\,=\,\sum_i\,\delta(\Vec{r}_i
        -\Vec{\tilde{r}}_1)
        \,\delta(\Vec{r}_i^\prime-\Vec{\tilde{r}}^\prime_1)\,\prod_{j\neq i}
        \,\delta(\Vec{r}_j-\Vec{r}^\prime_j)
\eeq
is the one-body density matrix operator; the primed quantities have
to  be evaluated at ${\bf r}_i'$ with $i=1, ...,A$. The integral of 
$n_D({\bf p}_m)$ gives the integrated nuclear transparency $T$ 
\begin{equation}
T = \frac{\int n_D({\bf p}_m)\, d{\bf p}_m}{\int n(k)\, d{\bf k}} = 
      \int \rho_D ({\bf r})\,d{\bf r} = 1+ \Delta T
\label{intnd}
\end{equation}
where $\rho_D({\bf r})=\rho_D ({\bf r}_1={\bf r}'_1\equiv {\bf r})$
and  $\Delta T$ originates from the FSI.
In Ref. \cite{cio01} Eq. (\ref{nd}) has been evaluated using a Glauber
representation for the scattering matrix $S$, \textit{viz}
\begin{equation}
S \rightarrow S_G({\bf r}_1\dots{\bf r}_A)=\prod_{j=2}^AG({\bf r}_1,{\bf r}_j)
      \equiv \prod_{j=2}^A\bigl[1-\theta(z_j-z_1)\Gamma({\bf b}_1-{\bf b}_j)\bigr]
\label{SG}
\end{equation}
where ${\bf b}_j$ and $z_j$ are the transverse and the longitudinal components 
of the nucleon coordinate ${\bf r}_j\equiv({\bf  b}_j,z_j)$, ${\mit\Gamma}({\bf b})$
the Glauber profile function for elastic proton nucleon scattering, and the 
function $\theta(z_j-z_1)$ takes care of the fact that the struck proton
``1''  propagates along a straight-path trajectory so that it interacts with
nucleon  ``$j$'' only if $z_j>z_1$. Generalizing the same cluster expansion described
in Section \ref{sectionCLUSTER} to take into account Glauber rescattering,
we have obtained the  \textit{distorted} nucleon momentum
distributions $n_D({\bf p}_m)=n_D(p_m, \theta)$, where $\theta$ is the angle
between ${\bf q}$ and ${\bf p}_m$; the results for $^{16}O$ and $^{40}Ca$
are presented in Fig. \ref{fig4}.
\begin{figure}[!hbp]
\centerline{\epsfig{file=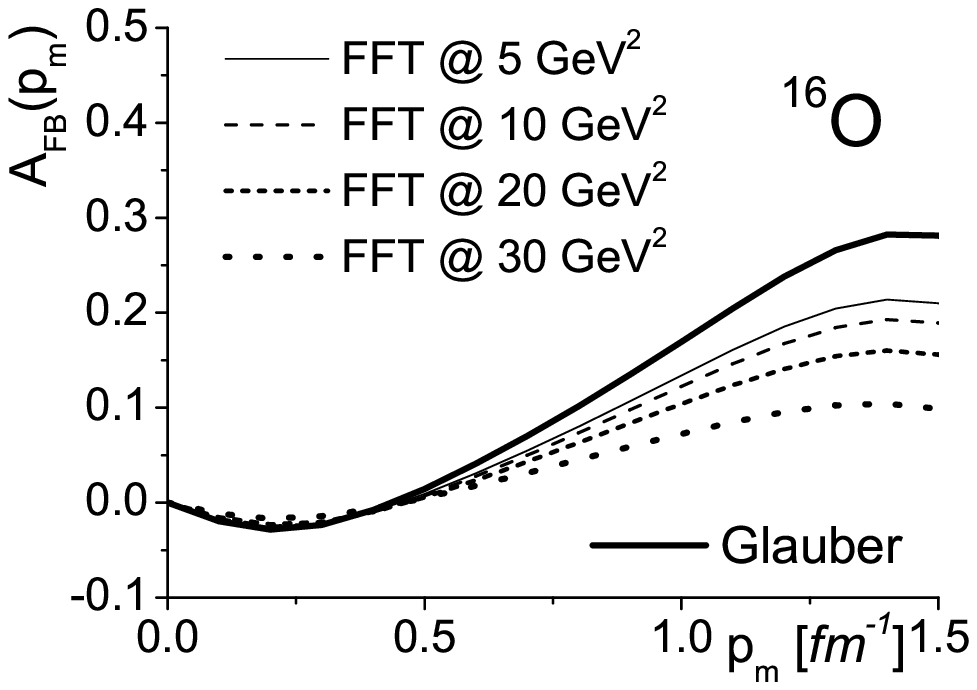,width=80mm}
   \epsfig{file=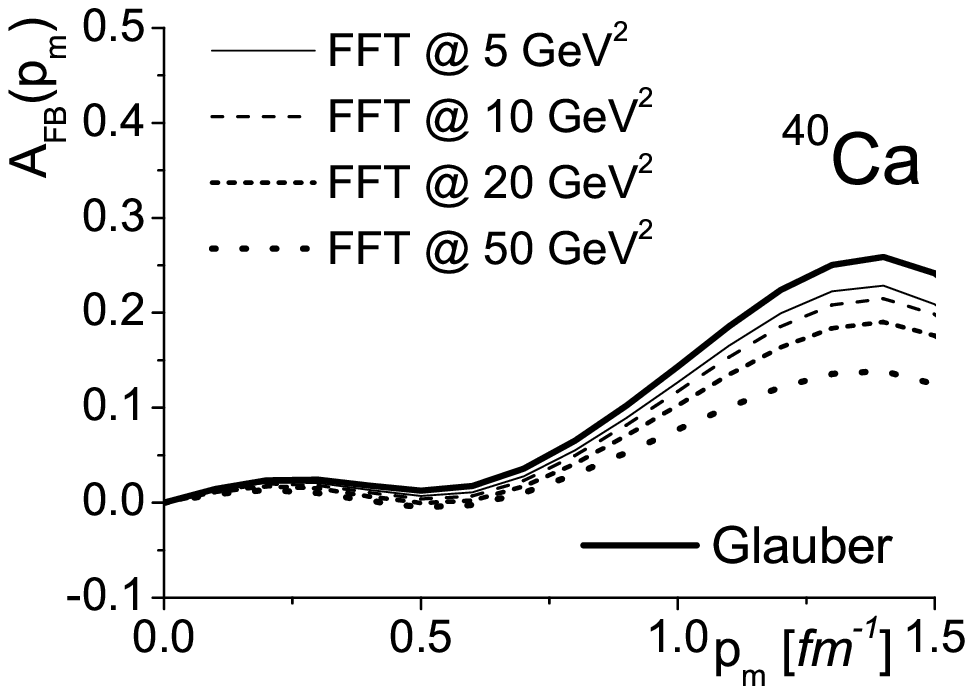,width=80mm}}
\caption{The forward-backward asymmetry (Eq. \ref{afb}) for the process
$A(e,e^\prime p)X$, with $A=16$ and $A=40$, calculated within the Glauber
approach (\textit{solid thick line}) taking into account the Finite Formation Time
(FFT) mechanism of various values of the 4-momentum transfer.}\label{fig5}
\end{figure}
The Glauber multiple scattering picture can be implemented by taking into
account Finite Formation Time (FFT) introduced in Ref. \cite{bra01}, where 
it has been shown that at the values of the Bjorken scaling variable
$x=Q^2/2m\nu \simeq 1$, FFT effects can be treated in a simple way, i.e. by 
replacing the Glauber operator (Eq. (\ref{SG})) with
\begin{equation}
\label{smatrix}
S_{FFT}({\bf r}_1,...,{\bf r}_A)=\,
      \,\prod^A_{j=2}\Big(1-J(z_1-z_j)\Gamma({\bf b}_1-{\bf b}_j)\Big)\,,
\end{equation}
where
\begin{equation}
J(z)\,=\,\theta(z)\,\Big(1\,-\,e^{-\,z\,\frac{x\,m\,M^2}{Q^2}}\Big)\,
\end{equation}
$m$ being  the nucleon mass  and $M^2=m^{*\,2}-m^2$ is a parameter 
describing the average excitation energy of the ejectile. It can be seen that 
at sufficiently high values of $Q^2$,  $J \rightarrow 0$ and the FSI vanishes;
this new mechanism for the description of FSI is inspired by the QCD prediction
of color transparency, which states that a pointlike, color-less particle, such as
the object singled out by the virtual photon interaction with the nuclear medium at
high $Q^2$, should have reduced cross-section with the nuclear medium as long as it
does not evolve into a physical proton inside the nucleus. We have calculated the
effect of FFT on the forward-backward asymmetry, defined in Ref. \cite{bia01} as
\beq
\label{afb}
A_{FB}\,=\,\frac{n_D(p_m,\theta=0^o)\,-\,n_D(p_m,\theta=180^o)}
      {n_D(p_m,\theta=0^o)\,+\,n_D(p_m,\theta=180^o)}
\eeq  
and the results are shown in Fig. \ref{fig5}.
\section{Applications. II - the total neutron-nucleus cross
  section}\label{sectionXSECT}

The neutron-Nucleus ($n-A$) total cross section, $\sigma^{tot}_{nA}$ is defined
by the \textit{optical theorem} as
\beq
\sigma_{tot}=\frac{4\pi}{k}Im\left[ F_{00}(0)\right]\,;
\eeq
where, within the Glauber eikonal approximation, the elastic scattering amplitude
has the following form:
\beq
F_{00}(\Vec{q})=\frac{ik}{2\pi}\int e^{i{\bf q}\cdot{\bf b_n}}
      \langle\psi_o|\{1-\hat{S}_G\}|\psi_o\rangle
=\frac{ik}{2\pi}\int d^2 b_n e^{i{\bf q}\cdot{\bf b_n}}\left[1 -
        \mathbf{e^{i\,\chi_{opt}({\bf b}_n)}}\right]
\eeq
with
\beq
\label{phsh}
\hspace{-1cm}\mathbf{e^{i\,\chi_{opt}(\Vec{b}_n)}}\,=\,\int
      \prod_{j=1}^A d\Vec{r}_j\,
      G(\Vec{b}_n,\Vec{s}_j)\,\left|\psi_o(\Vec{r}_1,...,\Vec{r}_A)\right|^2
      \delta\left(\frac{1}{A}\sum {\bf r}_j\right)\,,
\eeq
where $\Vec{r}_j\equiv(\Vec{s}_j,z_j)$ and $\Vec{b}_n$ is the neutron impact parameter.
As it is well known the squared wave function can be written in terms of density
distributions as follows \cite{alv04}:
\beq
\label{twenty8}
\left|\,\psi_o\,(\Vec{r}_1,...,\Vec{r}_A)\,\right|^2\,=\,\prod_{j=1}^A\,\rho(\Vec{r}_j)
      \,+\,\sum_{i<j}^A\,\mathbf{\Delta(\Vec{r}_i,\Vec{r}_j)}\,\prod_{k\neq(ij)}^A
      \,\rho(\Vec{r}_k)\,+\,...
\eeq
where $\rho(\Vec{r})$ is the one-body density distribution and the
\textit{two-body contraction}  $\Delta$ is defined in terms of one- and
two- body density distributions, $\rho^{(2)}(\Vec{r}_1,\Vec{r}_2)$:
\beq
\mathbf{\Delta(\Vec{r}_1,\Vec{r}_2)}\,=\,\left[\rho^{(2)}(\Vec{r}_1,
  \Vec{r}_2)\,-\,
  \rho^{(1)}(\Vec{r}_1)\,\rho^{(1)}(\Vec{r}_2)\right]\,;
\eeq
Usually Glauber-type calculations are based upon the single density approximation,
consisting in disregarding all terms of Eq. (\ref{twenty8}) but the first one. We
have considered, for the first time, also the second term of the expansion (\ref{twenty8})
\ie{ }the effects of two-body correlations. In the case of $^4He$ we have also considered
three- and for-body correlations \cite{alv04}; whereas for $^4He$ Eq. (\ref{twenty8}) has 
been used, for heavier nuclei we have used its optical limit ($A>>1$) in Eq. (\ref{phsh}) \ie
\beqy
\mathbf{e^{i\,\chi_{opt}(\Vec{b}_n)}}&\simeq&
    exp\left[-\,A\,\int d\Vec{r}_1\,\rho(\Vec{r}_1)\,
    \Gamma(\Vec{b}_n-\Vec{s}_1)\right.\,+\nonumber\\
&&\label{disti}+\,\left.A^2\,\frac{\int d\Vec{r}_1 d\Vec{r}_2\,
      \mathbf{\Delta(\Vec{r}_1,\Vec{r}_2)}\,
      \Gamma(\Vec{b}_n-\Vec{s}_1)\,\Gamma(\Vec{b}_n-\Vec{s}_2)}
      {1-\int d\Vec{r}_1\,\rho(\Vec{r}_1)\,\Gamma(\Vec{b}_n-\Vec{s}_1)}\right]\,;
\eeqy
which already for $A\ge 12$ reproduces the results based upon Eq. (\ref{twenty8})
almost exactly. We have evaluated the one- and two-body density matrices appearing
in Eq. (\ref{disti}) starting from the realistic wave functions obtained in Section 
\ref{sectionCLUSTER}.

The results of calculations are shown in Fig. \ref{fig7}. It can be seen that
the inclusion of correlations in the target wave function produce an enhancement
of the cross section of about $10\%$ with respect to the mean field result,
increasing the disagreement with the experimental data.
\begin{figure}[!htp]
\centerline{\epsfig{file=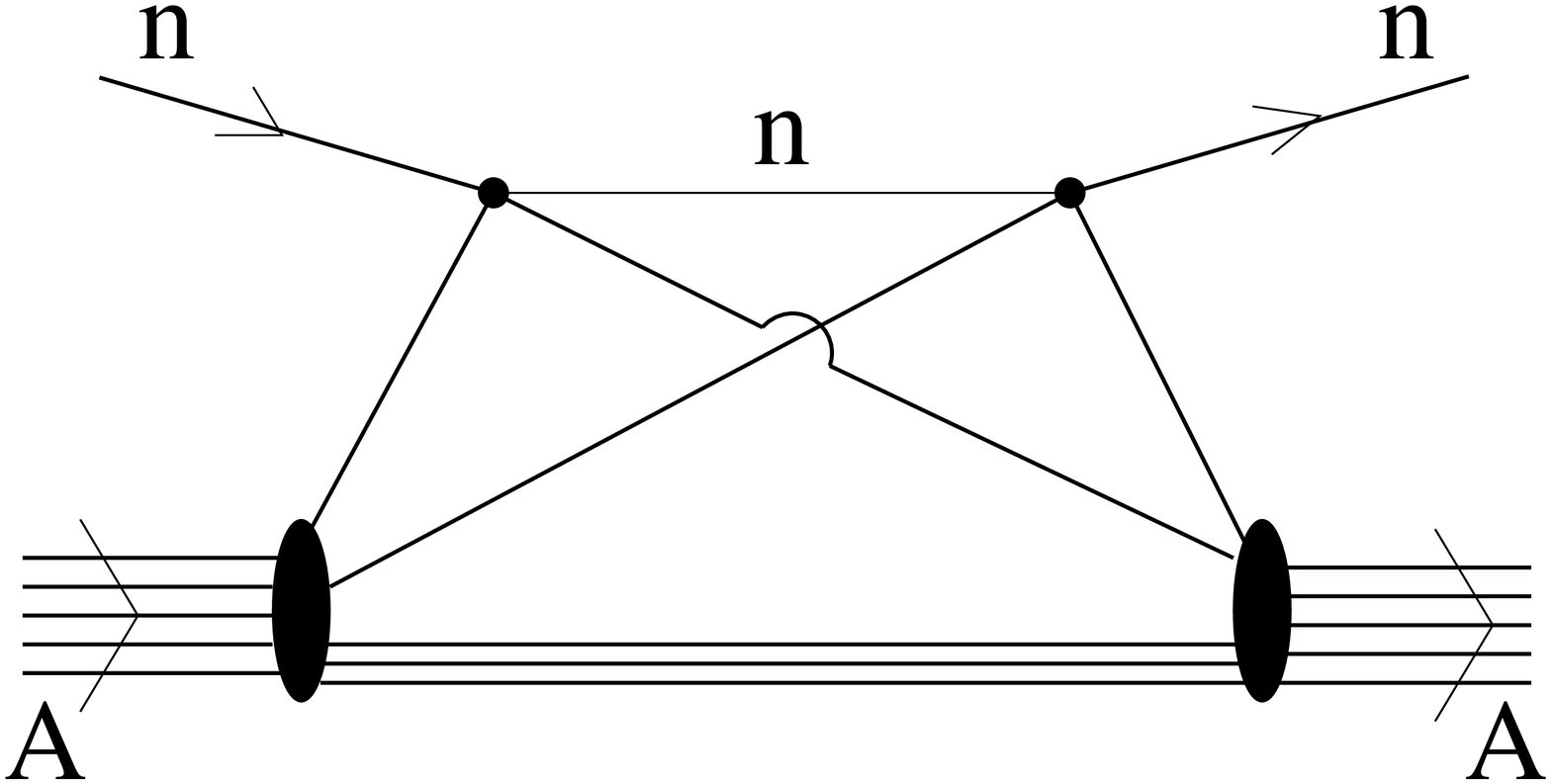,width=65mm}
   \hspace{0.5cm}\epsfig{file=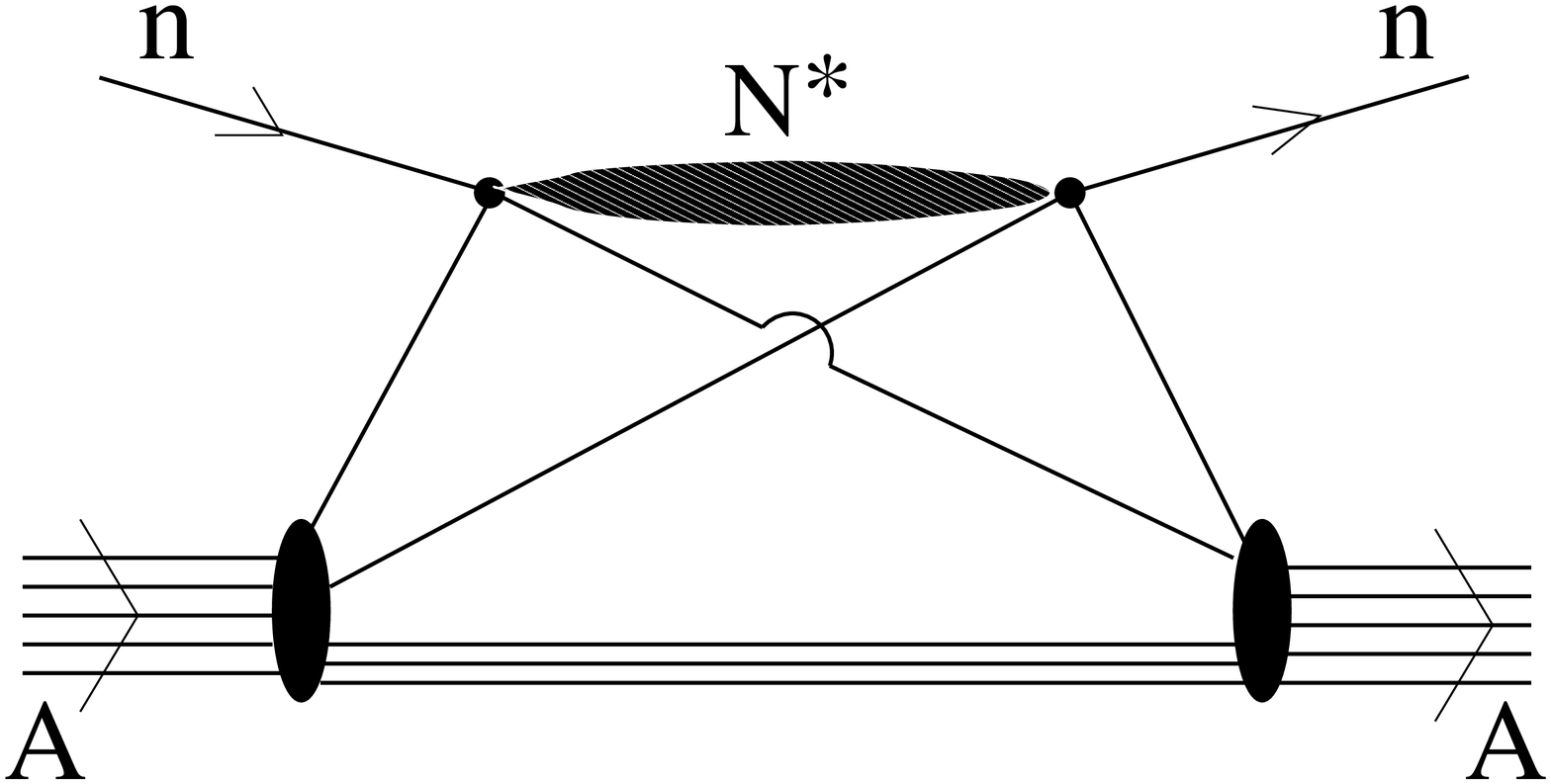,width=65mm}}
\centerline{\textit{a)}\hspace{5.5cm}\textit{b)}}
\caption{The neutron-nucleus total cross section diagrams. \textit{a)} Glauber double
scattering; \textit{b)} inelastic shadowing effects.}\label{fig6}
\end{figure}
It is well known however that at high energies diffractive scattering of the projectile,
depicted if Fig. \ref{fig6},
plays a relevant role. We have evaluated such an effect (also known as Gribov inelastic
shadowing) according to Refs. \cite{gri01,jen01}. It can be seen that inelastic shadowing
effects play indeed a relevant role to bring theoretical calculations 
in good agreement with experimental data.
\begin{figure}[!htp]
\centerline{\epsfig{file=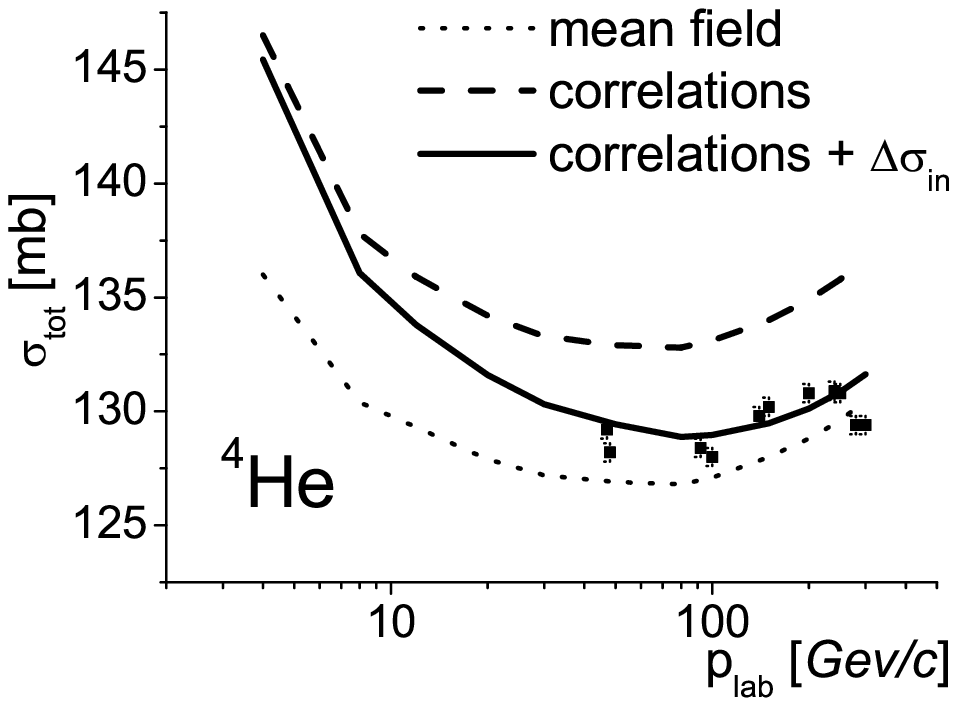,width=80mm}
   \epsfig{file=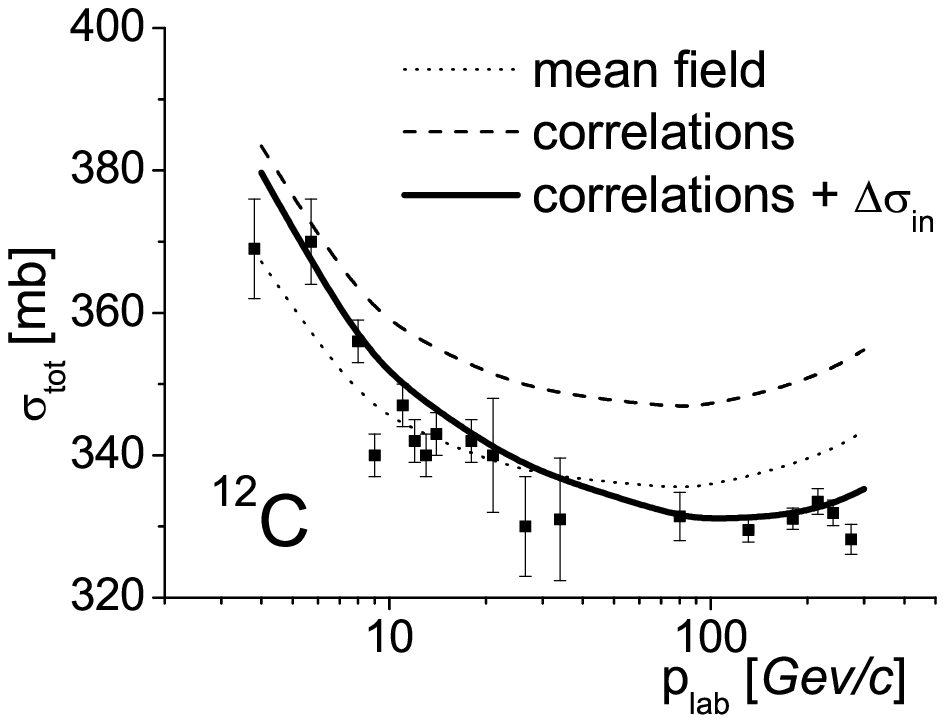,width=80mm}}
\centerline{\hspace{0.5cm}\epsfig{file=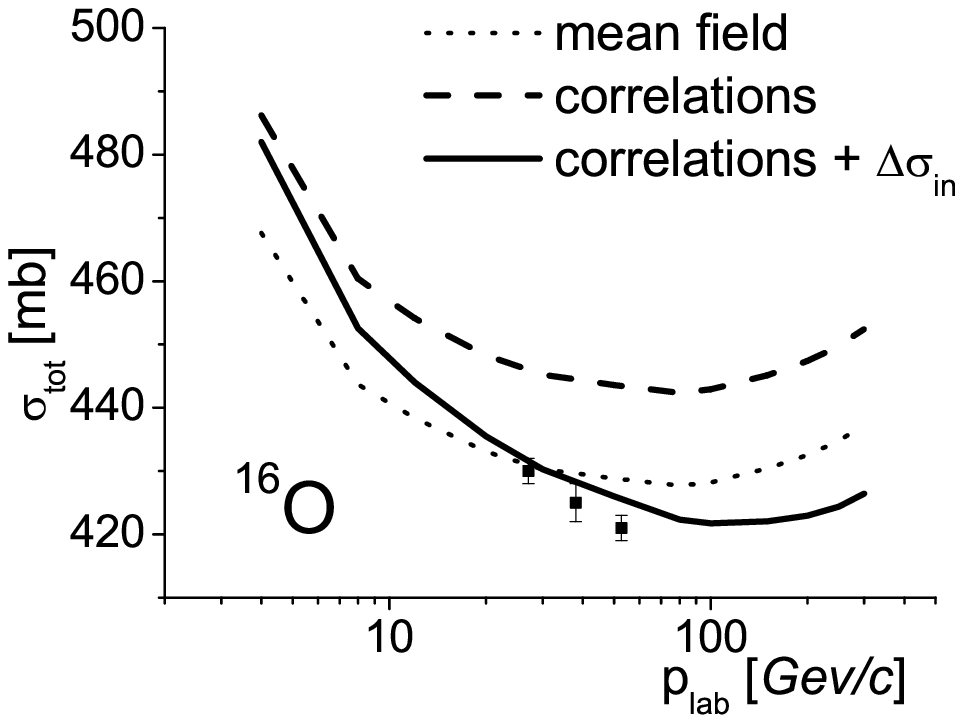,width=85mm,height=60mm}
   \hspace{-0.7cm}\epsfig{file=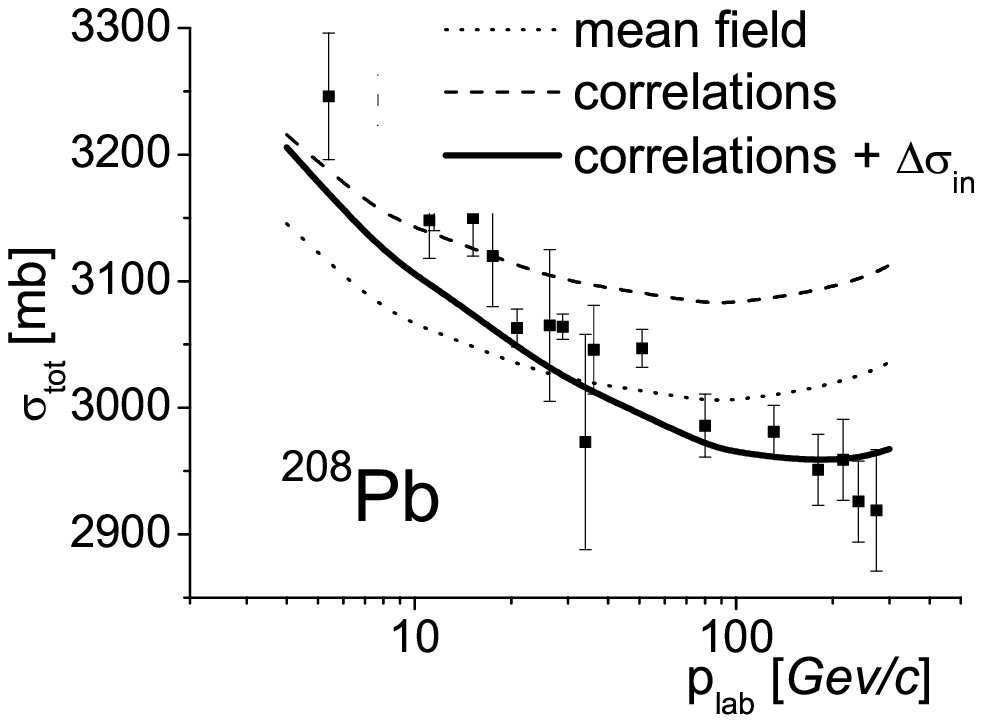,width=88mm,height=60mm}}
\caption{The double scattering diagrams in the neutron-nucleus total cross section.}
\label{fig7}
\end{figure}
\section{Summary}

We have developed  a method which
can be used to calculate scattering processes at medium and high energies
within a realistic and parameter-free  description of nuclear structure.
Our calculations followed  the following
strategy: \textit{i)} the values of the parameters pertaining to 
the correlation functions and
the mean field wave functions, have been  obtained from the calculation of the
ground-state
energy, radius and density of the nucleus,  to a given order of the expansion, using
realistic nucleon-nucleon interactions;
\textit{ii)} using these parameters we have calculated, by a proper generalization of the cluster
expansion, the distorted momentum distributions, the nuclear transparency and the 
total neutron-nucleus cross section.
 The method we have developed   appears to be a
very effective,  transparent and parameter-free one.
To sum up, we have shown that, using realistic models of the nucleon-nucleon interaction,
 a proper approach based on cluster expansion techniques  can produce reliable approximations
for those diagonal and non diagonal density matrices which appear in various
medium and high energy scattering processes off nuclei, so that the role of nuclear effects
in these processes can be reliably estimated without using free parameters to be fitted
to the data.

\section{Acknowledgments}

We are grateful to the organizers for the invitation to the Workshop.
 HM would like to 
thank the Department of Physics,
University of Perugia and INFN, Sezione di Perugia, for warm hospitality and support.
 Support by the Italian Ministero
dell'Istruzione, Universit\`a e Ricerca (MIUR), through the contracts
COFIN03-029498 and COFIN04-025729, is gratefully  acknowledged.


\end{document}